\documentclass[10pt,conference]{IEEEtran}

\usepackage{blindtext}
\usepackage{graphicx}
\usepackage{listings}
\usepackage{moresize}
\usepackage{url}
\usepackage{doi}
\usepackage[usenames, dvipsnames]{xcolor}

\makeatletter
\newcommand*\titleheader[1]{\gdef\@titleheader{#1}}
\AtBeginDocument{%
  \let\st@red@title\@title
  \def\@title{%
    \bgroup\normalfont\large\centering\@titleheader\par\egroup
    \vskip1.5em\st@red@title}
}
\makeatother

\title{Continuous Defect Prediction: The Idea and a Related Dataset\vspace{-10mm}}

\titleheader{\footnotesize{Lech Madeyski and Marcin Kawalerowicz. "Continuous Defect Prediction: The Idea and a Related Dataset". In: 14th International Conference on Mining Software Repositories. Buenos Aires. 2017, pp. 515--518. \doi{10.1109/MSR.2017.46}.  ~~~~~~~~~~~~~~~~~~~~~~~~~~~~~~~~~~~~~~~~~~~~~~~~~~~~~~~~~~~~~~~~~\href{http://madeyski.e-informatyka.pl/download/MadeyskiKawalerowiczMSR17.pdf}{\color{red}PDF}, \href{http://madeyski.e-informatyka.pl/download/MadeyskiRefs.bib}{\color{red}BibTeX}}\vspace{-14mm}}

\author{\authorblockN{Lech~Madeyski~\IEEEmembership{Member,~IEEE,}}
\authorblockA{Wroclaw University of Science and Technology,\\
Faculty of Computer Science and Management,\\
Wyb.Wyspianskiego 27, 50-370 Wroclaw, POLAND,\\
Lech.Madeyski@pwr.edu.pl}
\and
\authorblockN{Marcin~Kawalerowicz}
\authorblockA{Opole University of Technology, Faculty of Electrical\\
Engineering, Automatic Control and Informatics,\\
ul. Sosnkowskiego 31, 45-272 Opole, POLAND\\
and CODEFUSION Sp. z o.o., Armii Krajowej 16/2,\\
45-071 Opole, POLAND, marcin@kawalerowicz.net}
}

\begin{document}

\maketitle

\begin{abstract}
%\boldmath
%Note MK: Korektor zasugerowal: "We would like to present a dataset... " - nie podoba mi się
%We are sharing here a dataset that we created and use in our Continuous Defect Prediction (CDP) research. 
We would like to present the idea of our Continuous Defect Prediction (CDP) research and a related dataset that we created and share. Our dataset is currently a set of more than 11 million data rows, representing files involved in Continuous Integration (CI) builds, that synthesize the results of CI builds with data we mine from software repositories. Our dataset embraces 1265 software projects, 30,022 distinct commit authors and several software process metrics that in earlier research appeared to be useful in software defect prediction. In this particular dataset we use TravisTorrent as the source of CI data. TravisTorrent synthesizes commit level information from the Travis CI server and GitHub open-source projects repositories. We extend this data to a file change level and calculate the software process metrics that may be used, for example, as features to predict risky software changes that could break the build if committed to a repository with CI enabled.
%create integration build success/fail prediction models. 
\end{abstract}


\IEEEpeerreviewmaketitle
\bstctlcite{IEEEexample:BSTcontrol}

\section{Introduction}
Identifying defect-prone modules, packages or files has long intrigued researchers (e.g.,~\cite{Turhan09,DAmbros12,JureczkoMadeyski15}). However, limited resources and tight schedules common in software development environments have triggered a change of research focus into identifying defect-prone ("risky") software changes instead of files or packages. Our long-term research goal is to propose Continuous Defect Prediction (CDP) practice supported by a tool set using machine learning (ML)-based prediction models and large dataset (collected from both, open source and commercial projects) to predict defect-prone software changes (at the moment limited to success/fail continuous integration outcomes). We refer to our quality assurance practice as "Continuous Defect Prediction", as developers can receive a continuous feedback from the supporting tool we are working on as to whether the latest code change is risky and could break the build if committed/pushed/checked-in to the repository with Continuous Integration (CI) enabled.
We are building on top of previous works on classifying software changes either as being clean or buggy, e.g., of Kim et al.~\cite{kim2008}, Kamei et al.~\cite{kamei2013} and Yang et al.~\cite{yang2015}, as well as the TravisTorrent dataset made available by Beller at al.~\cite{msr17challenge}, but aim to deliver immediate and continuous feedback for a developer on how risky the most recent code change is, and to embed this feedback mechanism into the software development practice. Actually, this is a follow up to our previous research on Continuous Test-Driven Development~\cite{Madeyski13ENASE} and Agile Experimentation~\cite{MadeyskiKawalerowicz17}. Hence, we build upon our experience and tools developed so far.
In this paper, we share a large dataset used by our prediction models to identify defect-prone software changes.

Every build on a build server can be triggered by one or a set of commits (multiple commits pushed together to the central repository). First we are collecting information about those commits and then we calculate a set of metrics for every file change that took part in a given CI build. Those metrics include for example a number of modified lines (NML) since the last build, build commit local time of day (BCDTL), a number of distinct committers (NDC) involved in that build, a number of revisions (NR) for a given file, last build status (LBS) of the build where the file was involved. We use these metrics as features (predictors/independent variables) to create prediction models where the build result is the dependent variable. 

We are working on CDP in a commercial environment on a real life project where we use Jenkins CI build results from Bitbucket on-premise installation as a source of success/fail build indication. We combine this information with data collected from the Git repository. 
Due to the non-disclosure agreement we are able to share data collected from a large number of open source projects, but not from the commercial project we are involved in. 
We are sharing the data collected with help the of the TravisTorrent~\cite{msr17challenge} open database project. 
TravisTorrent synthesizes data from the Travis CI build server with the data collected from GHTorrent -- offline mirror of data provided by the API of popular GitHub version control hosting service. The data on TravisTorrent comes from popular open-source projects such as JRuby or Rails. TravisTorrent stores the data on a commit level meaning a commit is the most fine-grained piece of data it contains. In turn, we are working on a file change level. For us every file changed in a commit taking part in a CI build conveys meaningful information. By using this information, we build classification models and use them to predict the outcome of a build before the data is committed into the repository. Prediction models built on the basis of this dataset are beyond the scope of this data paper. 

We have collected more than 11 million rows of data for 1265 GitHub projects where more than 30,000 developers were active. 
%Now we 
We are making this data available for a broader public in hope it will help other researchers interested in defect-prone software change prediction, behaviour of software developers or other areas of software engineering research and practice.

\section{Data collection and storage}
We are collecting the CDP relevant data from two sources: continuous integration process and version control system (software repository).
The data from CI process are gathered from the CI server. We were interested in two pieces of information the CI server can provide us: build result and mark of the commit or commits involved in that build.
This information can be obtained from the CI server over an API or from another source, as an associated database, for example. The database usage can help to deal with temporary CI information. The data from the associated database is usually not cleaned as it is the case with the CI build information. It is customary to keep only the last $n$ builds information (e.g., date and time of the build, its result, build logs) to preserve the storage space on the CI server. We use TravisTorrent database to obtain the data on the build results and commits involved in those builds. TravisTorrent synthesizes the information taken from the Travis CI server using its API and GitHub repository data taken from its offline mirror (GHTorrent). We are specifically querying the \texttt{travistorrent\_11\_1\_2017} table and using the following columns:

\begin{enumerate}
\item git\_trigger\_commit (hash of the commit which triggered the build),
\item gh\_project\_name (project name on GitHub),
\item gh\_pushed\_at (time of the push that triggered the build),
\item tr\_status (build result).
\end{enumerate}

We are accessing Jenkins CI build results stored in a Bitbucket database in our CDP work on a commercial project. Both Jenkins and Bitbucket are installed on premise at a company building banking software. 
%Through loose coupling in our data gathering software we made it possible to use TeamCity CI server API to collect the build results from this system. 
Apart from Jenkins CI, we also made it possible to use the TeamCity CI server API to collect the build results. 
We support the Git source control system, however the approach does not limit us to this particular system by any means. It is possible to extend the approach to: Subversion, Mercurial or other source control systems. 

It is worth mentioning that depending on the chosen CI server, the build results enumeration can vary. We are mapping them to three states: success, failure, and unknown (for example if the build was interrupted, ended with a warning or was marked as unstable by the CI server), as shown in Table~\ref{tab:build_results}. 

\begin{table}[h]
\renewcommand{\arraystretch}{1.3}
\caption{Jenkins, Travis and TeamCity build results mapping}
\label{tab:build_results}
\centering
\begin{tabular}{|l|l|l|l|}
\hline
\textbf{Our database}& \textbf{Jenkins} & \textbf{Travis} & \textbf{TeamCity} \\ \hline
1 (success)          & SUCCESS    & passed  & NORMAL   \\ \hline
0 (failure)          & FAILURE    & failed  & FAILURE  \\ \hline
                     & NOT\_BUILT &         & ERROR    \\ \cline{2-2}\cline{4-4}
999 (unknown)        & ABORTED    & errored & WARNING  \\ \cline{2-2}\cline{4-4} 
                     & UNSTABLE   &        & UNKNOWN  \\ \hline
\end{tabular}
\end{table}

One build on a CI server can be made for one particular commit or for any number of distinct commits. The information on commits involved in a CI build can be obtained from a CI server API or from a database. 
We are querying the TravisTorrent database to get the information we need (using the \texttt{git\_trigger\_commit} column described earlier in this section) using the following query:

\begin{lstlisting}[frame=single,basicstyle=\ssmall\ttfamily,numbers=left,language=SQL]
SELECT git_trigger_commit, gh_project_name, tr_status, gh_pushed_at
FROM travistorrent_11_1_2017 
WHERE gh_project_name = :projectName 
AND gh_pushed_at IS NOT NULL
AND tr_build_id >= 
(SELECT tr_build_id FROM travistorrent_11_1_2017 
  WHERE git_trigger_commit=:commit_from)
AND tr_build_id <= 
(SELECT tr_build_id FROM travistorrent_11_1_2017
  WHERE git_trigger_commit=:commit_to)
ORDER BY tr_build_id DESC
\end{lstlisting}

The specific commits involved in the build are calculated based on the branch topology tree of the software repository.

Then the software repository is utilized as the source of the file level metrics gathered for CDP. Those metrics are the features for the prediction model. In that regard the type of the software repository (whether it is Git, Subversion, Mercurial or any other) from which the data were harvested is irrelevant. 
%Currently we are acquiring the data from Git repositories here GitHub to be specyfic. 
We are currently acquiring the data from Git repositories (GitHub to be specific). 
We are not using GitHub API to avoid problems with bandwidth throttling reported in~\cite{gousi13}, as in the case of TravisTorrent. We are cloning all of the GitHub repositories to a local disk instead. As a source of data we are using the commit history stored locally. As it turns out, this poses no problem to GitHub and is not a problem with regards to the space needed. We have cloned 1265 projects which occupied a disk space of little more than 43GB. To facilitate the Git repository operations (cloning, reading the commit history) we are using the LibGit2Sharp\footnote{\url{https://github.com/libgit2/libgit2sharp}} library.
%\footnote{https://github.com/libgit2/libgit2sharp} library. %\footnote{\url{https://github.com/libgit2/libgit2sharp}} library.

Table~\ref{tab:metrics} shows the metrics we are collecting from the software repository together with the information on to how they are acquired. The software process metrics we use were inspired by Madeyski and Jureczko~\cite{Madeyski15SQJ} who found that some process metrics (namely NDC and NML) can significantly improve software defect prediction models based on product metrics. 
Together with these metrics our dataset contains also:

\begin{enumerate}
    \item Project name
    \item File path
    \item Commit hash
    \item Build commit hash
    \item Export date (time stamp of the moment the data were collected)
\end{enumerate}

\begin{table}[h]
\renewcommand{\arraystretch}{1.3}
\caption{Metric harvested from the software repository}
\label{tab:metrics}
\centering
\begin{tabular}{|l|l|p{3.5cm}|}
\hline
\textbf{Metric}& \textbf{Abbr.} & \textbf{How acquired} \\ \hline
NumberOfRevisions & NR & Count the revisions participating in build\\ \hline
NumberOfDistinctCommiters & NDC & Count unique developers involved in revisions in build\\ \hline
NumberOfModifiedLines & NML & Count modified lines since the last build\\ \hline
NumberOfRevisions & NR & Count all the revisions of a given file\\ \hline
BuildDateTimeLocal & BDTL & Build server local date and time of the start of the build\\ \hline
BuildCommitDateTimeLocal & BCDTL & Local time stamp of the commit that caused the build\\ \hline
LastBuildStatus & LBS & Status of the previous build\\ \hline
AuthorIdentification & AI & Author Git user.name setting \\ \hline
\end{tabular}
\end{table}

The data are stored in a simple Microsoft SQL Server database. The part of the database that stores the metrics (table Metrics) is presented in Table~\ref{tab:database}. The rest of our database containing tables for defect prediction requests and classification models is not presented here as it is beyond the scope of this paper.

\begin{table}[h]
\renewcommand{\arraystretch}{1.3}
\caption{Microsoft SQL Server data storage column information for Metrics table}
\label{tab:database}
\centering
\begin{tabular}{|l|l|l|l|l|l}
\hline
\textbf{Column name}               & \textbf{Type} & \textbf{Nullable} \\ \hline
Id                         & bigint         & no \\ \hline
Path                       & nvarchar(500)  & yes \\ \hline
OldPath                    & nvarchar(500)  & yes \\ \hline
NumberOfRevisions          & int            & yes \\ \hline
NumberOfDistinctCommitters & int            & yes \\ \hline
NumberOfModifiedLines      & int            & yes \\ \hline
BuildResult                & int            & no \\ \hline
Commit                     & nvarchar(255)  & yes \\ \hline
BuildCommit                & nvarchar(255)  & yes \\ \hline
ExportDateUtc              & datetime       & no \\ \hline
NumberOfRevisions          & int            & yes \\ \hline
BuildDateTimeLocal         & datetime       & no \\ \hline
BuildCommitDateTimeLocal   & datetime       & no \\ \hline
BuildProjectName           & nvarchar(255)  & yes \\ \hline
Author                     & nvarchar(255)  & yes \\ \hline
PreviousBuildResult        & int            & yes \\ \hline
ProjectName                & nvarchar(255)  & yes \\ \hline
\end{tabular}
\end{table}

\section{Description of dataset}

\begin{figure}[h]
\centering
\includegraphics[width=3.5in]{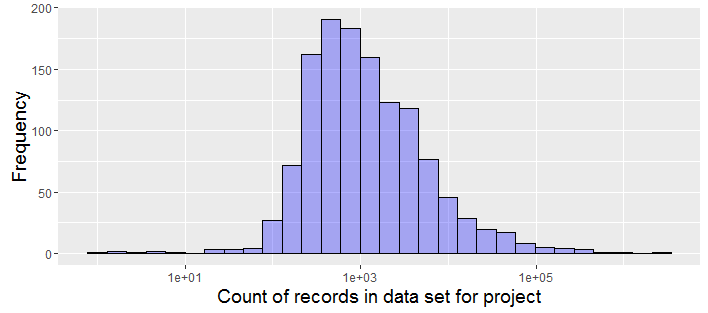}
\caption{Log10 scaled histogram of the count of file records per project in our database}
\label{fig:hist}
\end{figure}

At the time of writing this article we collected the dataset including 11,464,816 rows (representing files that participated in a CI build) in the Metrics database table. The metrics were collected for the 1265 projects gathered in the TravisTorrent database. The count of rows for different project vary drastically (with the minimum at 1, maximum at 2706617, 1st Quartile at 405, 3rd Quartile 2876, median at 938 and mean at 9063). Figure~\ref{fig:hist} shows the histogram of the rows count over the projects on a logarithmic scale. We have 30,022 distinct commit authors in our database. Interestingly there is a large number of projects with rather small number of data rows in our database, e.g., we have 657 projects with less than 1000 data records (1e+03 mark on Figure~\ref{fig:hist}), meaning all the commits in those projects changed less than 1000 files.  

\begin{figure*}[ht]
\centering
\includegraphics[width=6.7in]{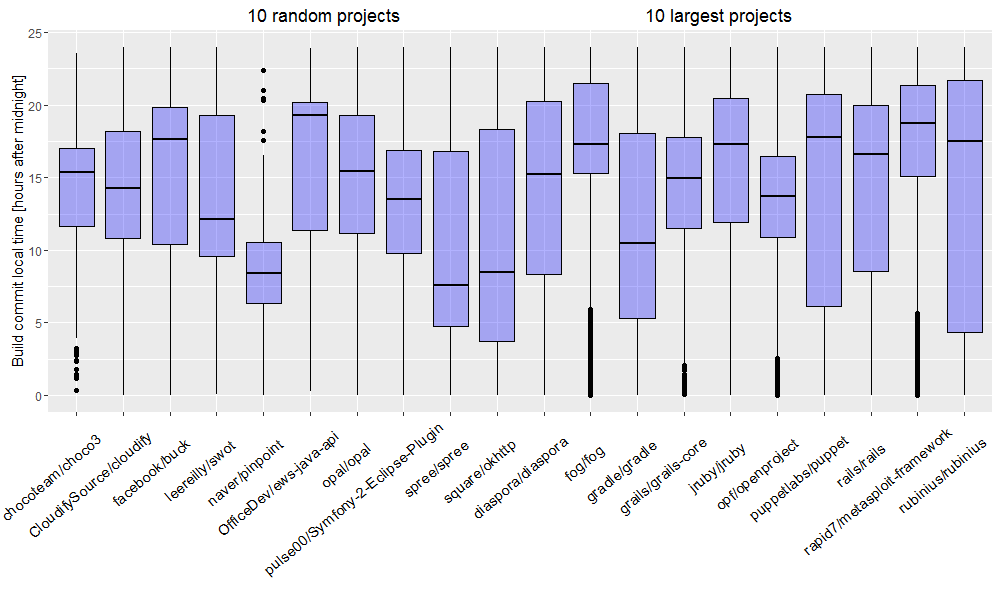}
\caption{Build commit local time for 10 random and 10 largest projects in our dataset}
\label{fig:btc}
\end{figure*}

We have done some exploratory data analysis for 10 largest projects in our database (the ones with the most rows) and for 10 randomly chosen (using the following SQL command: \texttt{select top 10 ... order by newid()} projects with rows count greater than the mean in our database.
% order by newid() daje w SQL Serwerze losową próbkę
%Table~\ref{tab:projects} shows those projects, while Table~\ref{tab:summary} presents the summary data for some of the metrics in those projects.
Figure~\ref{fig:btc} includes the names of the selected projects, while Table~\ref{tab:summary} presents the summary data for some of the metrics in those projects.

\begin{table}[!t]
\renewcommand{\arraystretch}{1.3}
\caption{Summary data for metrics for 10 random projects}
\label{tab:summary}
\centering
\begin{tabular}{|l|l|l|l|l|l|l|l|l|}
\hline
\textbf{}       & \multicolumn{2}{c|}{\textbf{NR}} & \multicolumn{2}{c|}{\textbf{NDC}} & \multicolumn{2}{c|}{\textbf{NML}} & \multicolumn{2}{c|}{\textbf{BCDTLl}} \\ \hline
\textbf{}       & \textbf{Rand}   & \textbf{Top}   & \textbf{Rand}    & \textbf{Top}   & \textbf{Rand}    & \textbf{Top}   & \textbf{Rand}     & \textbf{Top}     \\ \hline
\textbf{Min.}   & 1               & 1              & 1                & 1              & 0                & 0              & 0                 & 0                \\ \hline
\textbf{1stQu.} & 1               & 1              & 1                & 1              & 1                & 4              & 8.011             & 10.9             \\ \hline
\textbf{Med.}   & 1               & 1              & 1                & 1              & 4                & 20             & 13.01             & 16.7             \\ \hline
\textbf{Mean}   & 1.95            & 1.14           & 1.18             & 1.04           & 38.78            & 89.15          & 12.41             & 15.03            \\ \hline
\textbf{3rdQu.} & 1               & 1              & 1                & 1              & 26               & 68             & 18                & 20.03            \\ \hline
\textbf{Max.}   & 144             & 93             & 18               & 14             & 33718            & 98155          & 23.02             & 23.98            \\ \hline
\end{tabular}
\end{table}

During our CDP research, we work with both open source and closed source repositories. We share here only the open source based part of our dataset which poses a threat as it may not generalize to other contexts, e.g., commercial/closed source software projects. 
It is also worth mentioning that the TravisTorrent dataset we build upon restricted the project space using filtering criteria to Ruby or Java non-fork, non-toy, somewhat popular ($> 10$ watchers on GITHUB) projects with some history of TRAVIS CI use ($> 50$ builds)~\cite{msr17challenge}.

The goal of the paper is to briefly describe the dataset we have shared. However, our role as researchers, even if generally beyond the scope of this data paper, is not only to collect data, but also to transform them into understanding.
As an example of interesting insights or ideas for what future research questions could be answered with the provided dataset, we are presenting the box plots for the build commit times in the 10 largest large and 10 randomly chosen projects. Assuming these small samples are representative to some extent, we could draw two interesting, albeit preliminary, hypotheses from this plot. The first one, derived from Figure~\ref{fig:btc} and Table~\ref{tab:summary}, is that for the larger projects commits that trigger the CI build are done later in day time, whereas they are done earlier in the average. The second one is that the open source projects integrations are done generally well before 8 PM so on average open source developers are not night owls as they are usually perceived. Actually, the builds are done based on commits done in normal business hours, between 9 AM and 5 PM. 

\section{Future work and conclusions}
We are using the dataset presented in this paper to create prediction models for continuous prediction of success/fail continuous integrations. 
We are working currently on a pilot CDP project in a commercial software development environment, at a company managed by one of the authors of this paper, but we are going to release the whole CDP project together with tools used to gather the data using dual open source and commercial licences. 
We are going to enrich the dataset we collected with more metrics that, according to existing empirical evidence by other researchers, can be used as features in our prediction models. 
It might even be possible (through cooperation with one of the cloud providers) to expose our prediction models over the web in a ready to use manner to aid development of open source projects.

By opening our database to the public we hope to attract an audience and feedback to our project, as well as attract researchers to enhance the dataset by new metrics or new kinds of metrics, to build prediction models on the basis of a dataset from more than one thousand of software projects and even more unique developers, or to predict other dependent variables that can be useful for practitioners.
%Areas in which our data could aid future research are broad:
%\begin{itemize}
%\item defect prediction on software change level,
%\item stability and maturity studies on long running software projects,
%\item developers activity and results examination,
%\item exploration of trends in continuous integration over a period of time.
%\end{itemize}
Areas in which our data could aid future research are broad and include: defect prediction on software change level, stability and maturity studies on long running software projects, developers activity and results examination, or exploration of trends in continuous integration over a period of time.

%The database is available as a CSV (with semicolon as separator and double quote as string delimiter) file at \url{https://figshare.com/s/302814fa28cb1fbde705} or as a Microsoft SQL Server 2012 dump file at \url{https://figshare.com/s/394e2f8d7dc6da405721}. 
The database is available as a CSV (with semicolon as separator and double quote as string delimiter) file at figshare\footnote{\url{https://figshare.com/s/302814fa28cb1fbde705}} and as a Microsoft SQL Server 2012 dump file\footnote{\url{https://figshare.com/s/394e2f8d7dc6da405721}}. 
%The database is available as a CSV (with semicolon as separator and double quote as string delimiter) file at figshare\footnote{https://figshare.com/s/302814fa28cb1fbde705} and as a Microsoft SQL Server 2012 dump file\footnote{https://figshare.com/s/394e2f8d7dc6da405721}. 

\section*{Acknowledgment}

The authors would like to thank the authors of TravisTorrent~\cite{msr17challenge} who provided us a set of data that we were able to extend to use for our purposes. We would also like to thank the employees of the CODEFUSION company for their valuable input and help in tools development.

\ifCLASSOPTIONcaptionsoff
  \newpage
\fi

\ifCLASSOPTIONcaptionsoff
  \newpage
\fi

%\bibliographystyle{IEEEtran}

%\bibliography{IEEEabrv,main}


\end{document}